# Machine-learning classifiers for logographic name matching in public health applications: approaches for incorporating phonetic, visual, and keystroke similarity in large-scale probabilistic record linkage

Philip Collender[1], Zhiyue Tom Hu[1], Charles Li[1], Qu Cheng[1], Xintong Li[3], Yue You[1], Song Liang[4], Changhong Yang[2], Justin V. Remais[1*]

[1] University of California, Berkeley, Berkeley, California, USA

[2] Centers for Disease Control and Prevention, Chengdu, Sichuan, China

[3] Emory University, Atlanta, Georgia, USA

[4] University of Florida, Gainesville, Florida, USA

[*] Corresponding author: Justin V. Remais, PhD, Division of Environmental Health Sciences, School of Public Health, University of California, Berkeley, 2121 Berkeley Way #5302, Berkeley, CA 94720-7360; 510-643-8900 (tel); 510-643-5056 (fax); jvr@berkeley.edu.

**Abstract**

Approximate string-matching methods to account for complex variation in highly discriminatory text fields, such as personal names, can enhance probabilistic record linkage. However, discriminating between matching and non-matching strings is challenging for logographic scripts, where similarities in pronunciation, appearance, or keystroke sequence are not directly encoded in the string data. We leverage a large Chinese administrative dataset with known match status to develop logistic regression and Xgboost classifiers integrating measures of visual, phonetic, and keystroke similarity to enhance identification of potentially-matching name pairs. We evaluate three methods of leveraging name similarity scores in large-scale probabilistic record linkage, which can adapt to varying match prevalence and information in supporting fields: (1) setting a threshold score based on predicted quality of name-matching across all record pairs; (2) setting a threshold score based on predicted discriminatory power of the linkage model; and (3) using empirical score distributions among matches and nonmatches to perform Bayesian adjustment of matching probabilities estimated from exact-agreement linkage. In experiments on holdout data, as well as data simulated with varying name error rates and supporting fields, a logistic regression classifier incorporated via the Bayesian method demonstrated marked improvements over exact-agreement linkage with respect to discriminatory power, match probability estimation, and accuracy, reducing the total number of misclassified record pairs by 21% in test data and up to an average of 93% in simulated datasets. Our results demonstrate the value of incorporating visual, phonetic, and keystroke similarity for logographic name matching, as well as the promise of our Bayesian approach to leverage name-matching within large-scale record linkage.

# 1. Introduction

Record linkage, the task of identifying records corresponding to the same individual within or across databases, has many applications across the health sciences. As the number and availability of electronic health records and other relevant data grow, linkage across and within databases is ever more imperative to facilitate epidemiologic study of exposure-disease relationships (1), estimation of health registry completeness and true disease prevalence (2,3), and the study of health histories, co-morbidities, and co-infections (4–6). While many different approaches to record linkage exist, probabilistic record linkage, which utilizes statistical models to infer the probability that two records are co-referent based on agreement or similarity of identifying fields, offers advantages in terms of adapting to new data and estimating match uncertainty (7). Highly discriminatory identifiers, such as personal names, provide critical support to the linkage process but can be subject to complex variations in representation, typographical errors, and abbreviations, motivating the inclusion of a variety of subroutines to assess name identifier similarity during the linkage process (8,9).

For personal names, in order to accommodate variability that may arise through errors in the data entry process, as well as multiple ways of representing the same concept or entity (e.g., abbreviations, transliterations between languages, nicknames, reordering of given and family names), multiple heuristics have been developed to assess the similarity of character strings. Static approaches, such as cosine and Levenshtein similarities (10), impose a predefined penalty for each unit (character or multi-character token) of difference between two strings (8,11). In contrast, learnable similarity metrics can be trained using labeled examples of co-referent and non-co-referent names. Learnable similarity metrics encompass the use of machine learning classifiers trained on features extracted from pairwise string comparisons (such as static similarity scores) (12,13), as well as probabilistic finite-state transducers, which estimate the probability of specific variations in the context of surrounding characters (14–16). Training classifiers on multiple static similarity metrics can reveal non-linear relationships between similarity scores and matching probabilities and provide a means of empirically weighting various definitions of string similarity. Finite-state transducers, meanwhile, can learn penalties associated with specific character deletions, insertions, and substitutions in context, but are consequently much more complex, and require extensive training and validation data to achieve good performance.

Matching personal name variants based on string similarity is especially challenging for writing systems that do not use alphabetic scripts. Most method development for personal name-matching has targeted Indo-European languages, and particularly English, in which characters or combinations of characters correspond to phonemes, and the overall character set is relatively small. These properties allow for direct comparisons of character string data to serve as relatively good measures of similarity in terms of sound, appearance, and keystroke sequence (17). Under these conditions, static similarity scores applied directly to the original characters may exhibit favorable classification performance. In contrast, discriminating matching from non-matching name variants is much more challenging for logographic scripts, in which single characters encode words or concepts. Chinese scripts, for example, have over 50,000 distinct characters, many of which have highly similar or identical pronunciations, multiple possible pronunciations (17), visual appearances differing by a single stroke or radical, complex structures, such that composite characters are formed from multiple simple characters (18), and require multiple keystrokes to enter each character stored in a string. None of these complexities are directly encoded in representations of Chinese characters stored in computer memory, and thus traditional approaches to string similarity naively applied to Chinese character fields are unlikely to satisfactorily account for the errors in data recording and entry (Table 1), leading to poor ability to discriminate between co-referent and non-co-referent strings.

**Table 1:** Some observed variation types in Chinese personal names

| Error type | Example | Levenshtein similarity. |
|---|---|---|
| Character duplication | 万只子→万只只子 | 0.75 |
| Character transposition | 谯科江→谯江科 | 0.33 |
| Phonetic replacement | 张可成→张坷成<br>(zhang ke cheng → zhang ke cheng) | 0.67 |
| Visual replacement | 张雨雨→张币雨 | 0.5 |
| Radical decomposition | 阳娅→阳女亚 | 0.33 |

A partial solution to the challenges of assessing string similarity for logographic scripts is to encode the original logograms in formats that represent their phonetic or visual properties, or keystroke input sequences, before applying pairwise comparison methods, such as Levenshtein distance (10). A diverse set of encodings are available for this purpose: Pinyin, the official phonetic system for transcribing Chinese characters into the Latin alphabet, is also the most commonly used keyboard input method for Chinese PC users (19); Wubi, a keyboard input system preferred by some professional typists, is based on character structure, and can potentially represent aspects of visual and keystroke similarity (20); Four Corner Code is a numerical representation of stroke shapes around the outside edges of characters (21); characters may be further decomposed into their component radicals or strokes; and Unicode ideographic description characters can be used to indicate the structure of compound characters (e.g., , indicating a compound character composed of two radicals arranged horizontally; , indicating a compound character composed of two radicals arranged vertically).

However, because errors can occur in relation to numerous string properties, one may expect that multiple encodings and associated similarity measures, with some means of appropriately weighting their importance, may be required to approach optimal classification performance of co-referent names.

Here, we develop a framework leveraging multiple comparison features of encoded logographic names to recover aspects of phonetic, visual, and keystroke similarity using machine learning classifiers, with the goal of improving record linkage performance. We train gradient boosted classification tree ensembles and logistic regression classifiers on a large number of name pairs with known match status, identifying models that optimize tradeoffs between predictive performance and computational scalability. We investigate the impact of incorporating the resulting name-matching classifiers into large-scale record linkage using simulated data with varying rates of name errors and identifier availability, as well as real data withheld during model selection and calibration, proposing and evaluating three novel methods to either adapt name-matching thresholds to specific linkage tasks, or directly utilize continuous estimates of the probability of matching given a classifier score.

To our knowledge, ours is the first study to integrate measures of phonetic, visual, and keystroke string similarity for logographic name matching in the context of large-scale record linkage. Prior work has evaluated Chinese name-matching classifiers trained on pinyin similarity features to resolve co-referent name variations extracted from Chinese Wikipedia redirects (17), but did not evaluate the benefit of including other forms of similarity. Our approach addresses these methods gaps, and offers tools for researchers seeking to conduct epidemiologic analysis of large, linked datasets in logographic languages.

## 2. Methods

*2.1 Components of the name matching algorithm*
The main challenge to measuring string similarity for logographic languages is that the bytes encoding character data provide poor representations of pronunciation, visual structure, and keystroke sequences when compared with alphabetic character data. Thus, transformations of the original characters into encodings that more closely represent these properties are needed prior to the application of string comparators. Similarity measures along multiple dimensions may need to be subsequently combined into an overall similarity score. Our general approach is to use simple lookup tables to recover properties of logographic character strings, apply multiple pairwise similarity scoring algorithms to the resulting encoded character strings, and then feed the resulting scores as covariates to classification algorithms.

*String transformations*
To recover underlying properties relevant to the likelihood that two names are variants referring to the same entity, we map logographic character sequences to representations of their phonetic, visual, and input structures. We refer to these mappings as *string transformations*, and represent them using notation of the form $\mathcal{M}(S)$, where $S$ represents the original character string and $\mathcal{M}$ is replaced with a symbol representing the encoding. Encodings explored in the current study are described in **Table 2**. We note that, in addition to encodings related to phonetics, visual structure, and keystrokes, we extract information such as name frequency, likely ethnicity, and potential indicators of more than one name or name variant being present in a record. This information may be useful from a probabilistic standpoint (i.e., two common names are less likely to refer to the same person, even if highly similar; names of different ethnic origins may be unlikely to refer to the same person), or in terms of modifying the importance of various similarity scores (e.g., if phonetic mistakes are more common within one population group when compared with another).

Table 2: Encodings to represent phonetic, visual, keystroke inputs, and other properties of Chinese character data

| Encoding example | Description | Data source(s) |
|---|---|---|
| $\mathcal{I}(伍考) = 伍考$ | **Identity:** Direct comparison of logogram strings may be sufficient to exclude many non-matches, as well as to detect potential transposition errors | NA |
| $PY\left(伍考\right) = $ wu3_kao3 | **Pinyin:** The phonetic writing system for Chinese characters, as well as a commonly used keystroke input system. Related to *phonetic* and *keystroke* similarity. | https://github.com/haiwen/seahub/blob/master/seahub/convert-utf-8.txt |
| $FC\left(伍考\right) = $ 21212_44027 | **Four corner code**: A system for encoding Chinese characters based on the shapes found in the four corners of a character, as well as an extra digit describing the shape above the bottom right corner. Related to *visual* similarity. | http://bbs.unispim.com/forum.php?mod=viewthread&tid=31674 |
| $WB\left(伍考\right) = $ wgg_ftgn | **Wubi98**: A keyboard input system for entering Chinese characters based on their radical and stroke structures. Related to *visual* and *keystroke* similarity. | https://github.com/HesperusArcher/98WuBi |
| $RD\left(伍考\right) = $ 亻 一五ヰ一乙 | **Radical decomposition:** Breaking logograms down into their components. May be related to *visual* similarity, especially in cases where a character has been split into its subcomponents. | https://github.com/cjkvi/cjkvi-ids/blob/master/ids.txt |
| $RDS\left(伍考\right) = $ 亻 一五ヰ一乙⿰⿱ | **Radical decomposition with structural indicators:** Breaking logograms down into their component parts, and appending characters representing the structure of the original logograms. Related to *visual* similarity. | https://github.com/cjkvi/cjkvi-ids/blob/master/ids.txt |
| $HAN\left(伍考\right) = 1$ <br> $HAN(阿日么扎博) = 0$ | **Indicator for Han ethnicity:** We differentiate between names starting with one of the top 400 most popular Han Chinese surnames, and containing 2-4 logograms, and names not matching these criteria, since names following different ethnic conventions may be associated with different patterns of variation. | http://www.360doc.com/content/14/0713/00/906800_394004854.shtml |
| $AMB\left('俄者(拉者)'\right) = 1$ <br> $AMB\left('?者'\right) = 1$ | **Tally of indicators of ambiguity:** In some cases, name fields contain parenthetical variations on a name or character, question marks for unknown characters, or aliases. This encoding counts the number of instances of question marks, open and closed parentheses, and phrases such as 又名 (AKA), which may help adjust matching rules for such circumstances. | NA |
| $LF\left(伍考\right) = -9$ | **Log relative frequency of names:** High similarity between names that are common may be less indicative of a match than between rarer names. | NA |

*Pairwise comparisons of transformed strings*

Once logographic strings have been mapped to encodings that better represent their relevant properties, their pairwise similarities may be assessed using common scoring algorithms such as Levenshtein distance or cosine distance (10,22). String similarity functions differ in their penalization of various types of disagreement, with a particularly relevant distinction arising between methods such as Levenshtein distance, which are sensitive to the order of tokens (groups of consecutive characters) in both strings, and methods such as cosine distance, which consider only the set overlap of all tokens, and may therefore be more sensitive to detecting similarity in the case of character transpositions, but less specific in other cases. For each transformation relevant to phonetic, visual, or keystroke properties of logographic strings, a set of pairwise scoring methods was implemented based on Levenshtein similarity, cosine similarity (on tokens of length 1,2, or 3 characters), and longest-common-substring similarity, which we define here as the length of the longest sequence of shared characters divided by the length of the shortest string. Scoring functions were applied to substrings of the original input (e.g., 1st two logograms, 3rd-Nth (final) logograms) reasoning that different substrings, roughly corresponding to Han Chinese family and given names, may provide more or less discriminatory power to distinguish co-referent and non-co-referent name pairs, or may be subject to different types of error patterns. Additional properties, such as name frequency or presumed ethnicity, are summarized for each pair using simple functions. Resulting outputs are notated with the general form $\mathcal{M}_\mathcal{D}(\mathbf{S}, k, s_1:s_2)$, where $\mathcal{M}$ is a symbol representing the string transformation used, $\mathcal{D}$ is a symbol representing the comparison method, $k$ is the token length used, and $s_1:s_2$ is a range representing a contiguous subset of the original logograms, and summarized (for arbitrary $\mathcal{M}$) in **Table 3**.

Table 3: Pairwise comparison methods used to quantify Chinese name similarity

| Description | Relevant encodings | Token lengths[1] | Substring ranges[2] | # Features[3] |
|---|---|---|---|---|
| $LV$: **Levenshtein similarity** calculated as $1 - \frac{E}{\max(N_1,N_2)}$, where $E$ is the minimum number of character insertions, deletions, or swaps required to convert $S_1$ into $S_2$, $N_1$ and $N_2$ are their respective lengths | $\mathcal{I}$: Identity<br>$PY$: Pinyin<br>$FC$: Four corner code<br>$WB$: Wubi98<br>$RD$: Radical decomposition<br>$RDS$: Radical Decomposition w/ structure | 1 | 1:N, 1:1, 1:2, 2:N, 3:N | 30 |
| $LCS$: **Longest common substring** similarity, calculated as $\frac{\max(N_1,N_2)-E}{\min(N_1,N_2)}$ where $E$ is the minimum number of character insertions, deletions, or swaps required to convert $S_1$ into $S_2$, $N_1$ and $N_2$ are their respective lengths | $\mathcal{I}$: Identity<br>$PY$: Pinyin<br>$FC$: Four corner code<br>$WB$: Wubi98<br>$RD$: Radical decomposition<br>$RDS$: Radical Decomposition w/ structure | 1 | 1:N, 1:1, 1:2, 2:N, 3:N | 30 |
| $COS$: **Cosine similarity,** calculated as $\frac{\sum_i A_i B_i}{\sqrt{\sum_i A_i^2}\sqrt{\sum_i B_i^2}}$, for all tokens $i$ in $S_1$ or $S_2$, where $A_i$ is the frequency of token $i$ in $S_1$ and $B_i$ is its frequency in $S_2$ | $\mathcal{I}$: Identity<br>$PY$: Pinyin<br>$FC$: Four corner code<br>$WB$: Wubi98<br>$RD$: Radical decomposition<br>$RDS$: Radical Decomposition w/ structure | 1, 2, 3 | 1:N, 1:1, 1:2, 2:N, 3:N | 80 |
| $SUM$: Summing ambiguity indicator counts or log relative frequency (i.e., the overall log probability of comparing $S_1$ to $S_2$ among name pairs) for two strings | $AMB$: Ambiguity indicator counts<br>$LF$: Log relative frequency of name components | 1, 2, N-2, N-1, N | 1:N, 1:1, 1:2, 2:N, 3:N | 5 |
| $CAT$: Categorical encoding for Han name indicators, mapped to Both Han, Neither Han, or Disagreeing categories | $HAN$: Han name indicator | N | 1:N | 1 |

[1.] Token length refers to the number of consecutive characters of a transformed string that are considered as a single unit during string comparisons. Grouping characters can improve the specificity of similarity measures, especially those that do not intrinsically consider the order of characters in a string, such as cosine similarity.
[2.] Substring ranges describe the position of relevant logograms in the untransformed string, and may be roughly considered as representing comparisons of the full string (1:N), comparisons of positions corresponding to Han family names (1:1; 1:2), and comparisons of positions corresponding to Han given names (2:N, 3:N)
[3.] The number of features for each comparison method generally corresponds to # encodings × # token lengths × # substring ranges considered. Exceptions are that tokens of length 2-3 were not considered for cosine similarity of untransformed strings (the $\mathcal{J}$ encoding), summed counts of ambiguity indicators were considered only for full strings, and summed log relative frequencies were considered only for substring ranges 1:1, 1:2, 2:N, and 3:N

*Classification algorithms incorporating pairwise comparison output*
In order to integrate and appropriately weight various quantities related to phonetic, visual, and keystroke similarity of logographic name pairs, as well as other relevant properties, classification algorithms were trained to distinguish between name pairs of known match status $Y$ (1 if two names co-occur in records known to reference the same individual, 0 otherwise) based on J pairwise comparison features $\mathcal{M}_\mathcal{D}(\mathbf{S}, \mathrm{k}, \mathrm{s}_1{:}\mathrm{s}_2)_j$, yielding classification scores $X \approx P(Y = 1 | \mathcal{M}_\mathcal{D}(\mathbf{S}, \mathrm{k}, \mathrm{s}_1{:}\mathrm{s}_2)_1 \ldots \mathcal{M}_\mathcal{D}(\mathbf{S}, \mathrm{k}, \mathrm{s}_1{:}\mathrm{s}_2)_J)$. While many learning algorithms could be leveraged for this purpose, three are considered here: (1) single comparison features used directly as classification scores; (2) multiple logistic regression; and (3) boosted classification tree ensembles as implemented in the *Xgboost* software package (*R* package version 0.81.0.1). These three approaches cover a range of tradeoffs in expected bias and variance, as well as in computational complexity, which is a potentially crucial consideration for large-scale linkages where billions or trillions of record pairs must be evaluated. Single comparison scores are expected to provide limited distinction between co-referent and non-co-referent name pairs, but may provide adequate improvements to performance when computational resources are limited. Logistic regression relies on an assumption of a linear relationship between the log-odds of co-reference and similarity scores, which may be a source of bias, on the one hand, yet may help to reduce out-of-sample variance by limiting the potential for overfitting. Logistic regression also provides a simple platform to specify and test the predictive value of interaction terms, facilitating incorporation of domain knowledge. Boosted classification tree ensembles are the most data-driven and computationally demanding of the approaches, capable of fitting non-linear relationships and discovering interactions between terms. However, they have a higher potential for overfitting to training and validation data, and therefore potentially more variance in their out-of-sample performance.

*Classification performance*
To evaluate the performance of the classifiers, two metrics of their ability to rank co-referent name pairs higher than non-co-referent pairs were jointly considered: Area under the Receiver-Operator Characteristic curve (AUROC), and area under the Early Recovery Portion of the Receiver-Operator Characteristic curve (EAUROC). EAUROC is defined as the area under the curve up to a false positive rate equal to the overall odds of true positive labels in the dataset. The motivation for considering both metrics stems from the extreme class imbalance evident in most record linkage applications, under which matching record pairs are so rare that only a very small false positive rate may conceivably yield an acceptable recovery in terms of precision. AUROC is indicative of ranking performance for the full dataset, and may be more relevant as class imbalance is reduced. EAUROC is indicative of ranking performance in the region of the ROC curve where the number of false positives is less than or equal to the total number of true positives.

*2.2 Training name matching algorithms*

All data processing and computation were carried out in the *R* software environment for statistical computing (version 3.5.2). Model selection and hyperparameter tuning for name-matching classifiers were undertaken using routines available in the *mlr* package (v 2.13).

*Test data and data partitioning*
Name matching algorithms were trained, tuned, and tested on data from an administrative dataset originating from Sichuan Province, China with known links (982,350 records representing 432,446 co-referent clusters). In order to tune the predictive capabilities of our name matching classifiers and test their out-of-sample performance, records were partitioned into training, development, and test sets in an approximately 60/20/20 split. During the splitting process, we avoided breaking linkages between any co-referent pairs of names (i.e., distinct names assigned to the same cluster). As a result, 593,201 records (261,033 clusters, 203,005 unique names) were assigned to the training partition, 194,811 records (85,816 clusters, 67,515 unique names) to the development partition, and 195,938 records (86,306 clusters, 67,465 unique names) to the test partition.

Co-referent and non-co-referent name pairs were sampled for each partition in order to train, tune, and test logistic regression and Xgboost classifiers. We sampled all non-identical co-referent name pairs within each partition (7,156 in the training data, 2,216 in the development data, 2,169 in the test data), along with large random samples of non-co-referent name pairs ($2 \times 10^6, 1 \times 10^6, 1 \times 10^6$), and generated pairwise comparison features for the selected name pairs. For the purposes of name-matching, all name pairs that ever occurred in co-referent record clusters were labeled as matches. The institutional review board committee at the University of California, Berkeley, approved all study protocols.

*Logistic regression model selection*
The optimal logistic regression classifiers were determined using a two-step process. In the first step, forward variable selection on all single features was used to identify the best-performing predictive model according to AUROC and EAUROC metrics on the development data, stopping once improvement in both metrics fell below $1 \times 10^{-5}$. In the second step, all variables initially selected were included as both main effects and interacted with categorical indicators for both names in a pair corresponding to Han Chinese naming conventions, neither name corresponding to Han conventions, or disagreement between the apparent ethnicity of the names. Starting from this full interaction model, terms were sequentially dropped until AUROC or EAUROC worsened by more than $1 \times 10^{-5}$.

*Xgboost tuning and model selection*
In order to reduce the risk of overfitting by the Xgboost algorithm, the functional relationships between pairwise similarity scores and the probability of co-reference were *a priori* constrained to be monotonic increasing. Tuning and model selection was performed in three steps: (1) an initial round of hyperparameter tuning was conducted on critical parameters such as maximum tree depth, learning rate, and the number of boosted trees; (2) sequential feature ablation was performed, removing at each iteration the feature with lowest permutation importance as measured by deterioration in out-of-sample EAUROC when the values of that column were permuted (23); and (3) upon identifying the best-performing feature set identified during sequential ablation, a final round of hyperparameter tuning was performed, optimizing simultaneously for out-of-sample EAUROC, AUROC, and minimum prediction time. Hyperparameter tuning was performed using Bayesian model-based optimization, available through the R package *mlrMBO* (v. 1.1.2).

2.3 *Applying name-matching classifiers within large-scale record linkage*

*Record linkage system*
The highly computationally efficient open-source probabilistic record linkage package *fastLink* (v. 0.5.1) (7) was adapted to utilize the name-matching classifiers described here during record linkage. The package follows an extended Fellegi-Sunter framework (24) for probabilistic record linkage. Briefly, patterns of full or partial agreement on identifying fields are tallied across all record pairs, resulting in an observed matrix $\boldsymbol{\gamma}$ of $J$ observed agreement patterns across $F$ fields of the form

$$\begin{matrix} \gamma_{1,1} & \cdots & \gamma_{1,F} \\ 0 & \cdots & 1 \\ 1 & \cdots & 1 \\ \cdots & \cdots & \cdots \\ \gamma_{J,1} & \cdots & \gamma_{J,F} \end{matrix}$$

with associated frequency counts $N_j$. The observed agreement patterns arise from a mixture of matching and non-matching pairs, and, following initialization of parameters $\pi_m$, the overall proportion of matching record pairs, $p(\gamma_j|M)$, the conditional probability of observing agreement pattern $j$ among matched record pairs, and $p(\gamma_j|U)$, the conditional probability of observing the pattern among non-matched record pairs, a mixture likelihood is optimized via EM algorithm. Upon convergence, the probability of matching associated with each agreement pattern $\gamma_j$, $\zeta_j$, is estimated according to Bayes' Law:

$$\zeta_j = \frac{\pi_m p(\gamma_j|M)}{\pi_m p(\gamma_j|M) + (1-\pi_m)p(\gamma_j|U)}$$

In addition to highly efficient computation and storage of agreement information, *fastLink* incorporates extensions of the Fellegi-Sunter framework to include conditional dependence between fields via interaction terms in log-linear models, as well as for inferring the match probability of agreement patterns with missing data under the assumption that missingness is random conditional on match status.

*Incorporating name-matching classifiers into probabilistic record linkage*
A typical approach to incorporating approximate name-matching into probabilistic record linkage has been to apply a threshold to similarity scores, above which all name pairs are declared to be in agreement. However, little guidance exists in the literature with respect to sound methods for selecting a threshold score. The optimal agreement threshold for any given linkage problem can be expected to depend on the overall proportion of matching record pairs, as well as the discriminatory power provided by other fields. Therefore, we evaluated two methods for selecting a classifier score threshold for agreement based on initial estimates of $\boldsymbol{\zeta_j}$ obtained by defining agreement on each field via exact matching, and utilizing the empirical distribution of classifier scores among matching and nonmatching record pairs observed in the development linkage dataset (18,975,565,455 record pairs). In the first approach, a threshold is selected to maximize the predicted $F_1$ score (harmonic mean of recall and precision) of name agreement among all record pairs that do not match exactly on name. Let $j_0$ denote the rows of $\boldsymbol{\gamma}$ for which $\gamma_{j,name} = 0$. The estimated recall at threshold $\tau$ is $P(X \geq \tau|M)$, based on the empirical distribution, while estimated precision is given by the average of

$$\frac{\zeta_j P(X \geq \tau|M)}{\zeta_j P(X \geq \tau|M) + (1-\zeta_j) P(X \geq \tau|U)}$$

across all $j \in j_0$, weighted by their corresponding $N_j$. We refer to this method of agreement threshold selection as $\tau_1$. The second proposed approach for estimating the optimal agreement threshold, referred to as $\tau_2$, is based on the theoretical best-possible ranking resulting from the predicted transfer of matching and non-matching record pairs from rows of $\boldsymbol{\gamma_j}$ where $\gamma_{j,name} = 0$ to corresponding rows where $\gamma_{j,name} = 1$, and all other agreement indicators are identical. For each donor row $j1$ and recipient row $j2$, the predicted $\zeta_j$ and $N_j$ following approximate name-matching at threshold $\tau$ are given by

$$\widehat{\zeta_{j1}} = \frac{\zeta_{j1}(1 - P(X \geq \tau|M))}{\zeta_{j1}(1 - P(X \geq \tau|M)) + (1-\zeta_{j1})(1 - P(X \geq \tau|U))}$$

$$\widehat{\zeta_{j2}} = \frac{\zeta_{j2} N_{j2} + \zeta_{j1}(1 - P(X \geq \tau|M)) N_{j1}}{N_{j2} + N_{j1}\left(\zeta_{j1} P(X \geq \tau|M) + (1-\zeta_{j1}) P(X \geq \tau|U)\right)}$$

$$\widehat{N_{j1}} = N_{j1}\left(\zeta_{j1}(1 - P(X \geq \tau|M)) + (1-\zeta_{j1})(1 - P(X \geq \tau|U))\right)$$

$$\widehat{N_{j2}} = N_{j2} + N_{j1}\left(\zeta_{j1} P(X \geq \tau|M) + (1-\zeta_{j1}) P(X \geq \tau|U)\right)$$

Following calculation of these predicted quantities at all relevant rows of $\boldsymbol{\gamma_j}$, rows are sorted by decreasing order of $\hat{\zeta}_j$, and AUROC is calculated. For both approaches, we assessed the optimal classifier score threshold based on enumeration of predicted performance over a regular 10,000 point grid of candidate values.

In addition to the two threshold-based methods of incorporating name-matching classifiers into the record linkage framework discussed above, we assessed the performance of a third method, which retains name-matching classification score information at a greater resolution. In this approach, which we label posterior name-matching adjustment, the matching probabilities of record pairs pertaining to selected rows of $\gamma_j$ for which $\gamma_{j,Name} = 0$ are re-estimated by using preliminary estimates of $\zeta_j$ as prior probabilities, along with smoothed estimates of the relative likelihood of observed name classifier scores based on empirical distributions in the linkage development data. Under this framework, the new estimate of $\zeta_j$ given $X$ is

$$\widehat{\zeta_j} = \frac{\zeta_j \hat{f}(X|M)}{\zeta_j \hat{f}(X|M) + (1-\zeta_j)\, \hat{f}(X|U)}$$

where $f(X|M)$ and $f(X|U)$ are empirical probability densities for matches and nonmatches at score $X$. In order to ensure desirable behavior, predictions are generated from isotonic regressions, which enforce the constraint that the predicted probability of matching increases monotonically with classifier score. In addition to retaining more granular representations of name similarity, this method has the potential advantage of ignoring large numbers of record pairs for which preliminary estimates of $\zeta_j$ are so low that no matches are expected to be recoverable (e.g., no posterior probabilities above 0.1 are possible). As a result, it may be faster and more scalable than the threshold-based methods. On the other hand, the ultimate reliance on name-similarity scores may not always be robust in comparison with EM algorithm estimates based on patterns of agreement over multiple fields.

*Record linkage application and evaluation*

The performance of record linkage methods incorporating logographic name-matching classifiers was tested on an administrative dataset from Sichuan Province, China, apportioned to development (18,975,565,455 record pairs, true match probability $7.74 \times 10^{-6}$) and test data (19,195,751,953 record pairs, true match probability $7.64 \times 10^{-6}$). Record linkage was conducted considering name, sex, year, month, and day of birth, and a location code (LOC code) specifying the district of residence for the record. In order to mitigate violations of the baseline assumption of conditional independence in agreement between fields (e.g., if there is a higher likelihood of random agreement on name given agreement on sex, or a higher likelihood of errors in data entry for month of birth given errors in day of birth), we selected from a set of all pairwise interaction terms for $p(\gamma_j|M)$ and $p(\gamma_j|U)$, as well as three-way interaction terms selected on the basis of known patterns of error in the data or hypothesized trends in naming practice. These included a three-way interaction for year, month, and day of birth, since all were derived from the same original field; between LOC code, month, and day of birth, under the assumption that observed patterns of over-attribution to particular dates (e.g., the 1st of the current month) may be localized to certain reporting sites; between name, sex, and LOC code, under the assumption that there may be regional trends in popular names for each gender; and name, sex, and year of birth, under the assumption that there may be temporal trends in popular names for each gender. Selection of the best-performing set of terms for matches and non-matches was performed using a genetic algorithm optimizing on AUROC and mean absolute error of estimate match probabilities on the development data (package *GA*, v. 3.2). Following identification of an adequate model, we evaluated linkage based on exact agreement, as well as incorporating the best-performing single pairwise similarity score, logistic regression classifier, and Xgboost classifier via the $\tau_1$, $\tau_2$, and posterior name-matching adjustment methods. We evaluated linkage performance based on AUROC, EAUROC, negative log-likelihood (-LL) of the estimated probabilities given true match status, as well as the number of false negatives (FN) and false positives (FP) when accepting the top-ranked agreement patterns as links up to a threshold minimizing the absolute difference between the true ($\pi_m$) or estimated ($\hat{\pi}_m$) proportion of matches and the proportion of record pairs accepted as matches.

In the interest of assessing the average impact of our methods under varying rates of error in the name field and availability of other identifiers, we generated 100 simulated paired datasets of 10,000 records to be linked under 5%, 10%, or 20% error in the name field. Data were generated based on marginal rates of error in sex, year, month, day of birth, and LOC code observed in the full set of labeled records, and with the same number of unique values and pairwise correspondence between distinct values as observed in the real data, but with no conditional dependence between errors in each field. Names were simulated by randomly sampling from the observed distribution of characters in each position of the real names, including a 'STOP' character to control name length. Distributions of types of name errors (e.g., single character replacement, character insertion, character transposition, character decomposition, or complex errors) were simulated in accordance with the observed data, with candidate character substitutions and decompositions determined by a combination of observed examples and additional substitutions having a Levenshtein similarity score of at least 0.8 according to one or more encodings. We evaluated performance of record linkage based on exact field agreement, as well as incorporating the best-performing single pairwise similarity score, logistic regression classifier, and Xgboost classifier via the $\tau_1$, $\tau_2$, and posterior name-matching adjustment methods for each dataset when all identifying fields were available, when only name, sex, year, month, and day of birth were available, and when only name, sex, and year of birth were available.

## 3. Results

*3.1 Distribution of errors in administrative data*
Among the labeled administrative records, 3.45% of co-referent records disagree on name. Among 25,680 co-referent record pairs that do not agree exactly on name, 92.56% of disagreements were due to replacement of one (84.48%) or more characters, 2.87% were due to character insertions or deletions, 2.36% due to character transpositions, 0.59% due to inclusion of multiple or alternative names/characters in the field, 0.50% due to decomposition of single into multiple characters, and 1.11% due to complex errors incorporating multiple types of variation. With respect to the independent discriminative power of each field for record linkage, sensitivities of exact agreement for each field are 96.55% for name, 99.47% for sex, 98.22% for year of birth, 96.33% for month of birth, 95.35% for day of birth, and 88.22% for LOC code. Specificities are >99.99% for name, 46.59% for sex, 98.18% for year of birth, 91.64% for month of birth, 96.64% for day of birth, and 99.93% for LOC code. Thus, probabilistic linkage based on exact field agreement is expected to exhibit high baseline performance in this dataset.

*3.2 Selected name matching classifiers and variable importance*
The comparison features included in the pareto-optimal set of single similarity scores were all cosine similarities on tokens of length 2-3, and included the pinyin and Wubi 98 string transformations. Similarity of strings encoded in pinyin and Wubi 98 may confer advantages in low dimensional classification settings, since both encodings are potentially related to keystroke inputs, as well as phonetic and structural similarity, respectively.

The best-performing logistic regression classifier included nine comparison features with effects stratified by ethnic categorization of the name pairs (i.e., neither Han, both Han, or one Han and the other not; Table 4). The model included a diversity of comparison features, with the greatest change in log odds of co-reference associated with increased Levenshtein similarity of pinyin encodings for non-Han and disparately categorized name pairs, while the largest effect among Han name pairs was associated with increased cosine similarity of trigram four-corner code. In general, features associated with visual and structural similarity based on four corner, Wubi 98, and radical decomposition + structure features were associated with greater increases in the odds of matching among name pairs classified as both Han.

Table 4: Change in log odds of co-reference associated with observed range of features in best logistic regression classifier

| Feature | Han name pairs | Non-Han name pairs | Name pairs with disagreeing ethnic categorization |
|---|---|---|---|
| Intercept | -21.14 | -14.65 | -14.65 |
| $PY_{LV}(\boldsymbol{S}, 1, 1{:}N)$ | 4.51 | 7.84 | 8.81 |
| $FC_{COS}(\boldsymbol{S}, 3, 1{:}N)$ | 7.81 | 5.29 | 6.34 |
| $WB_{LV}(\boldsymbol{S}, 1, 1{:}N)$ | 6.92 | 2.05 | 1.17 |
| $PY_{COS}(\boldsymbol{S}, 3, 1{:}N)$ | 5.93 | 4.28 | 5.93 |
| $WB_{COS}(\boldsymbol{S}, 3, 1{:}N)$ | -0.86 | 3.68 | 3.68 |
| $RDS_{COS}(\boldsymbol{S}, 2, 1{:}N)$ | 3.30 | 1.38 | 1.38 |
| $LF_{SUM}(\boldsymbol{S}, 2, 1{:}2)$ | -3.29 | -1.88 | 0 |
| $WB_{COS}(\boldsymbol{S}, 1, 1{:}2)$ | -0.39 | -1.34 | -1.57 |
| $RDS_{COS}(\boldsymbol{S}, 3, 1{:}2)$ | 0.64 | 0.54 | 1.02 |

The best-performing Xgboost classifier included 34 features, covering a wide variety of string transformations and comparison functions. Among the top 10 features ranked by permutation importance (Figure 1, left panel), four are summed log relative frequencies of different substrings of the names, which were also the highest-ranked features in terms of the frequency with which they were included in classification trees (Figure 1c, right panel), but were not directly associated with large information gains (Figure 1b, center panel), which may suggest a high degree of interaction with other features. Interestingly, the largest direct information gains were associated with Levenshtein and cosine similarities of untransformed strings, which may have to do with a large proportion of non-matching name pairs having no logograms in common, and with instances of logogram transposition, respectively.

Figure 1: Variable importance among top 10 features for best performing Xgboost classifier

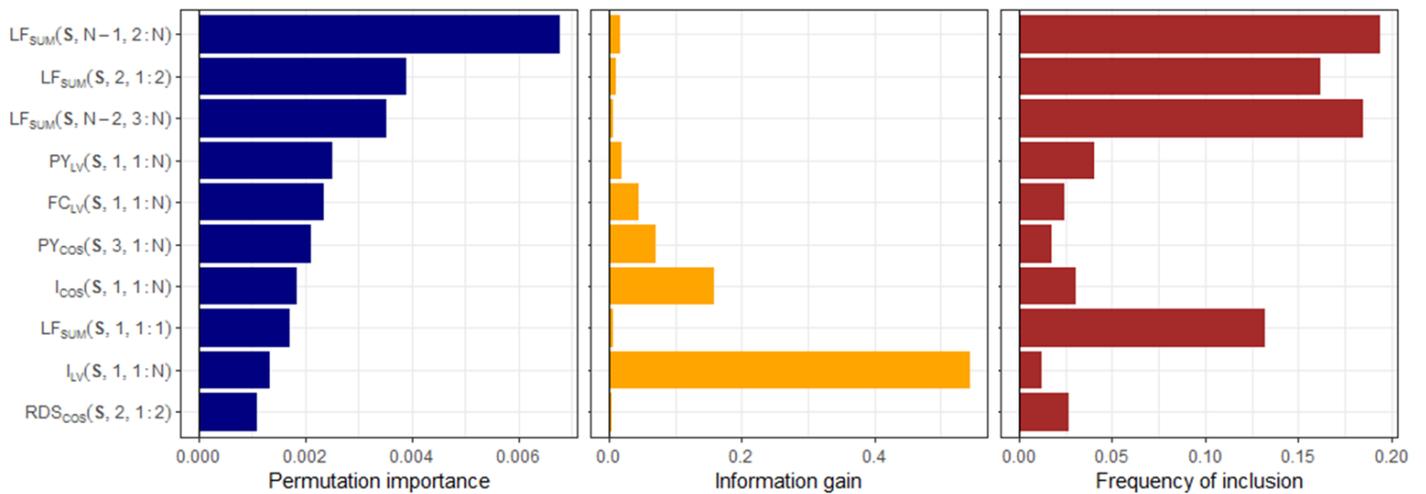

The pareto-optimal set of best-performing pairwise similarity measures, as well as selected logistic regression and Xgboost classifiers all exhibited strong ranking performance, with AUROCs consistently above 0.99 for full record linkage datasets in the development, test, and simulated data (Table 5). As expected, both logistic regression and Xgboost classifiers outperformed single similarity measures, especially with respect to early recovery of matching names. While the selected Xgboost classifier appeared to slightly out-perform logistic regression within the development name-matching data, the logistic regression classifier consistently exhibited better early recovery on full record linkage datasets, perhaps reflecting both the generally simple relationships between similarity scores and matching probabilities and the tendency for linear models to exhibit better generalizability from limited training data (while our name-matching datasets contain a large number of name pairs, they are highly imbalanced and the amount of matching examples for particular types of variation may be small) (25).

Table 5: Performance of pareto-optimal single pairwise similarity scores, logistic regression, and Xgboost name-matching classifiers within development name-matching data and across all record pairs in development, test, and simulated data

|  | Name matching data | Full record linkage data | | |
|---|---|---|---|---|
| **Classifier** | Development AUROC (EAUROC[1]) | Development AUROC (EAUROC[2]) | Test AUROC (EAUROC[2]) | Simulated data AUROC (EAUROC[2]) |
| $\mathcal{WB}_{cos}(\mathbf{S}, 3, 1:N)$ | 0.9976 (0.9194) | 0.9976 (0.0177) | 0.9963 (0.0125) | 0.9951 (0.0099) |
| $\mathcal{WB}_{cos}(\mathbf{S}, 2, 1:N)$ | 0.9978 (0.8676) | 0.9976 (0.0199) | 0.9953 (0.0179) | 0.9945 (0.0201) |
| $\mathcal{PY}_{cos}(\mathbf{S}, 3, 1:N)$ | 0.9978 (0.8341) | 0.9977 (0.0062) | 0.9972 (0.0069) | 0.9953 (0.0093) |
| Logistic regression | 0.9995 (0.9733) | 0.9995 (0.1467) | 0.9989 (0.1328) | 0.9992 (0.1188) |
| Xgboost | 0.9996 (0.9797) | 0.9995 (0.0934) | 0.9991 (0.1295) | 0.9993 (0.0984) |

[1.] Area under early recovery region of receiving operator characteristic curve up to a false positive rate equal to the overall odds of matching in the data. For name-matching data, this corresponds to a maximum FPR of $2.22 \times 10^{-3}$
[2.] For ease of comparison, all EAUROCs within record linkage datasets are referenced to a maximum FPR of $7.74 \times 10^{-6}$, which is the odds of finding a matching record pair in the development record linkage dataset

### 3.3 Linkage of development and test data

Based on model selection via genetic algorithm, pairwise terms for conditional dependence in agreement of name and year of birth, sex and year of birth, sex and month of birth, year of birth and day of birth, and year of birth and month of birth were included in the model for $p(\gamma_j|M)$, while the model for $p(\gamma_j|U)$ included interactions between agreements on name and sex, sex and LOC code, sex and day of birth, LOC code and month of birth, year of birth and day of birth, as well as a three way interaction between year, month, and day of birth.

Application of the various linkage methods incorporating name-matching classifiers to development and test data revealed that proposed methods always improved on linkage using exact agreement on names in terms of early ranking performance, log likelihood of estimated match probabilities given true data labels, and misclassification rates when accepting approximately the top true $\pi_m$ proportion of record pairs, ranked by estimated match probability (Tables 6 and 7). There was no clear trend in the performance of name-matching classifiers across all methods, although the logistic regression and Xgboost classifiers consistently outperformed trigram pinyin cosine similarity in terms of early ranking and log likelihood when incorporated using the $\tau_2$ and posterior adjustment methods, which more aggressively leverage discriminatory support provided by other fields. Furthermore, there was a general trend towards improved ranking, log likelihood, and misclassification rates when accepting the top $\pi_m$ proportion of record pairs, improving from $\tau_1$ incorporation to $\tau_2$ incorporation to posterior adjustment for all classifiers. On the other hand, misclassification rates did not improve as reliably when accepting approximately the estimated top $\hat{\pi}_m$ proportion of record pairs, suggesting that the more aggressive $\tau_2$ and posterior adjustment methods of performing linkage with fuzzy name matching may have a tendency to over-estimate the proportion of matched pairs in the data. False positive rates when accepting approximately the highest ranked $\hat{\pi}_m$ proportion of record pairs were substantially higher for the test data than for the development data when using logistic regression or Xgboost incorporated via $\tau_2$, and for all classifiers when using posterior adjustment, suggesting that $\hat{\pi}_m$ was substantially overestimated by these methods in the test data. This degradation in performance could result from either extrapolation of the underlying agreement correlation structure optimized for development data, or from differences in the distribution of name-match classifier scores among co-referent and non-co-referent record pairs between the two partitions.

Table 6: Linkage performance on development data by name-matching classifier and method of incorporation

| Classifier | Incorporation method | EAUROC[1] | -LL[2] | $FN_{\pi.m}$[3] | $FP_{\pi.m}$[4] | $FN_{\hat{\pi}.m}$[5] | $FP_{\hat{\pi}.m}$[6] |
|---|---|---|---|---|---|---|---|
| Exact agreement | - | 0.980513 | 44,692 | 4,671 | 2,396 | 3,301 | 4,135 |
| | | ΔEAUROC | Δ-LL | $\Delta FN_{\pi.m}$ | $\Delta FP_{\pi.m}$ | $\Delta FN_{\hat{\pi}.m}$ | $\Delta FP_{\hat{\pi}.m}$ |
| $PY_{COS}(\mathbf{S}, 3, 1:N)$ | $\tau_1$ | 0.002874 | -4,758 | -384 | 34 | -449 | 146 |
| Logistic regression | $\tau_1$ | 0.003565 | -6,913 | -367 | -39 | -427 | -124 |
| Xgboost | $\tau_1$ | 0.001683 | -2,670 | -216 | 9 | -258 | 30 |
| $PY_{COS}(\mathbf{S}, 3, 1:N)$ | $\tau_2$ | 0.003394 | -5,621 | -467 | 34 | -541 | 201 |
| Logistic regression | $\tau_2$ | 0.006663 | -8,893 | 233 | -838 | -1007 | 856 |
| Xgboost | $\tau_2$ | 0.006378 | -9,271 | -1,217 | 718 | -1,012 | 221 |
| $PY_{COS}(\mathbf{S}, 3, 1:N)$ | posterior adjustment | 0.008658 | -9,646 | -1,852 | 366 | -912 | -78 |
| Logistic regression | posterior adjustment | 0.009836 | -12,001 | -886 | -740 | -1,241 | 121 |
| Xgboost | posterior adjustment | 0.009870 | -11,795 | -963 | -682 | -1,298 | 246 |

[1]EAUROC: area under early recovery portion of ROC curve, up to a maximum false positive rate of $\frac{\pi_m}{1-\pi_m}$;
[2]-LL: Negative log likelihood (or log loss) of estimated match probabilities
[3]$FN_{\pi.m}$: Number of false negatives when accepting approximately the top true $\pi_m$ record pairs, ranked by estimated match probability
[4]$FP_{\pi.m}$: Number of false positives when accepting approximately the top true $\pi_m$ record pairs, ranked by estimated match probability;
[5]$FN_{\hat{\pi}.m}$: Number of false negatives when accepting approximately the top estimated $\hat{\pi}_m$ record pairs, ranked by estimated match probability;
[6]$FP_{\hat{\pi}.m}$: Number of false positives when accepting approximately the top estimated $\hat{\pi}_m$ record pairs, ranked by estimated match probability

Table 7: Linkage performance on test data by name-matching classifier and method of incorporation

| Classifier | Incorporation method | EAUROC[1] | -LL[2] | $FN_{\pi.m}$[3] | $FP_{\pi.m}$[4] | $FN_{\hat{\pi}.m}$[5] | $FP_{\hat{\pi}.m}$[6] |
|---|---|---|---|---|---|---|---|
| Exact agreement | - | 0.979204 | 46,517 | 4,838 | 2,525 | 3,483 | 3,746 |
| | | ΔEAUROC | Δ-LL | $\Delta FN_{\pi.m}$ | $\Delta FP_{\pi.m}$ | $\Delta FN_{\hat{\pi}.m}$ | $\Delta FP_{\hat{\pi}.m}$ |
| $PY_{COS}(\mathbf{S}, 3, 1:N)$ | $\tau_1$ | 0.004631 | -4,192 | -422 | 38 | -471 | 87 |
| Logistic regression | $\tau_1$ | 0.003346 | -6,355 | -338 | -78 | -382 | -149 |
| Xgboost | $\tau_1$ | 0.001449 | -2,117 | -191 | 4 | -216 | 9 |
| $PY_{COS}(\mathbf{S}, 3, 1:N)$ | $\tau_2$ | 0.003217 | -4,713 | -479 | 40 | -535 | 122 |
| Logistic regression | $\tau_2$ | 0.005996 | -7,590 | -1,044 | 758 | -971 | 1,321 |
| Xgboost | $\tau_2$ | 0.006130 | -8,588 | -1,431 | 851 | -1,114 | 1,680 |
| $PY_{COS}(\mathbf{S}, 3, 1:N)$ | posterior adjustment | 0.008792 | -8,952 | -1,798 | 442 | -991 | 854 |
| Logistic regression | posterior adjustment | 0.009841 | -10,225 | -954 | -573 | -1,332 | 1,140 |
| Xgboost | posterior adjustment | 0.009781 | -10,379 | -1,096 | -450 | -1,364 | 1,498 |

[1]EAUROC: area under early recovery portion of ROC curve, up to a maximum false positive rate of $\frac{\pi_m}{1-\pi_m}$;

[2]-LL: Negative log likelihood (or log loss) of estimated match probabilities

[3]$FN_{\pi.m}$: Number of false negatives when accepting approximately the top true $\pi_m$ record pairs, ranked by estimated match probability

[4]$FP_{\pi.m}$: Number of false positives when accepting approximately the top true $\pi_m$ record pairs, ranked by estimated match probability;

[5]$FN_{\hat{\pi}.m}$: Number of false negatives when accepting approximately the top estimated $\hat{\pi}_m$ record pairs, ranked by estimated match probability;

[6]$FP_{\hat{\pi}.m}$: Number of false positives when accepting approximately the top estimated $\hat{\pi}_m$ record pairs, ranked by estimated match probability

*3.4 Linkage results on simulated datasets*
When name, sex, year, month, and day of birth, and LOC code were available as identifiers, linkage performance on simulated data generally demonstrated that incorporation of name-matching classifiers resulted in improvements to ranking, probability estimation, and recovery when thresholding at the top-ranked true or estimated proportion of matching record pairs (Figure 2). Logistic regression and Xgboost name-matching classifiers tended to outperform name-matching using a single pairwise similarity score, $PY_{COS}(\mathbf{S}, 3, 1:N)$, and performance using the same classifier tended to be best using the posterior adjustment method of incorporation, followed by the $\tau_2$ thresholding method, followed by $\tau_1$. Gains in performance were more dramatic under more prevalent name errors among matching record pairs. For example, the mean total number of misclassifications when accepting the top estimated $\hat{\pi}_m$ proportion of record pairs fell by 88.43% from 121 (95% CI: 92-151) to 14 (95% CI: 7-24) when using the logistic regression classifier via posterior adjustment among datasets with name error rates of 5%, while among datasets with name error rates of 20%, it fell by 92.64% from 421 (95% CI: 382-469) to 31 (95% CI: 21-40). Decreases in both the number of false positives and false negatives at the estimated $\hat{\pi}_m$ threshold suggest that incorporation of name-matching classifiers generally improved the estimation of overall match probability in this setting.

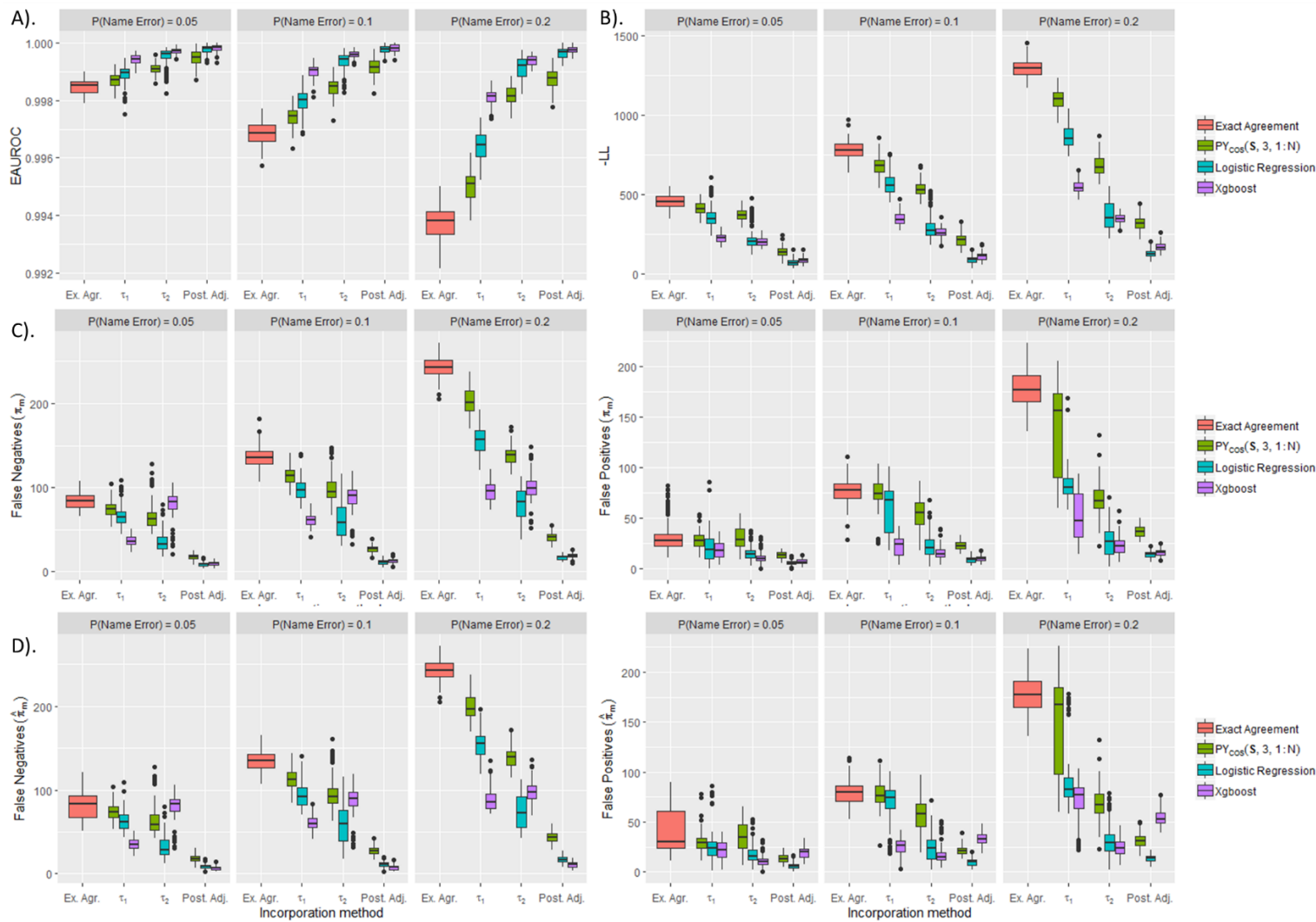

Figure 2: Linkage performance on simulated datasets with 5%, 10%, and 20% name error rates, when name, sex, year, month, and day of birth, and LOC code are available using various name-matching classifiers and methods of incorporation. Results are grouped by A). ranking performance (area under early portion of ROC curve); B). quality of estimated match probabilities (negative log likelihood given true match distribution); C). misclassifications when accepting the top ranked $\pi_m$ proportion of record pairs; D). misclassifications when accepting the estimated top ranked $\hat{\pi}_m$ proportion of record pairs. Each box plot corresponds to 100 simulations.

When LOC code was not available as an identifier, similar improvements to ranking and probability estimation were generally evident after incorporation of name-matching classifiers, quality of recovery when thresholding at the top-ranked true or estimated proportion of matching record pairs was more variable, with increasing numbers of false positives when implementing name-matching methods (Figure 3). This is due in part to superior baseline exclusion of false positives by exact-agreement linkage in this setting, although a trend towards increased false positive rates is evident for the $\tau_2$ and posterior adjustment methods of name-match classifier incorporation. In particular, some instability in the quality of recovery under the $\tau_2$ thresholding method appears to emerge in this setting, suggesting that it may be particularly sensitive to noise in the initial estimates of match probabilities $\zeta_j$.

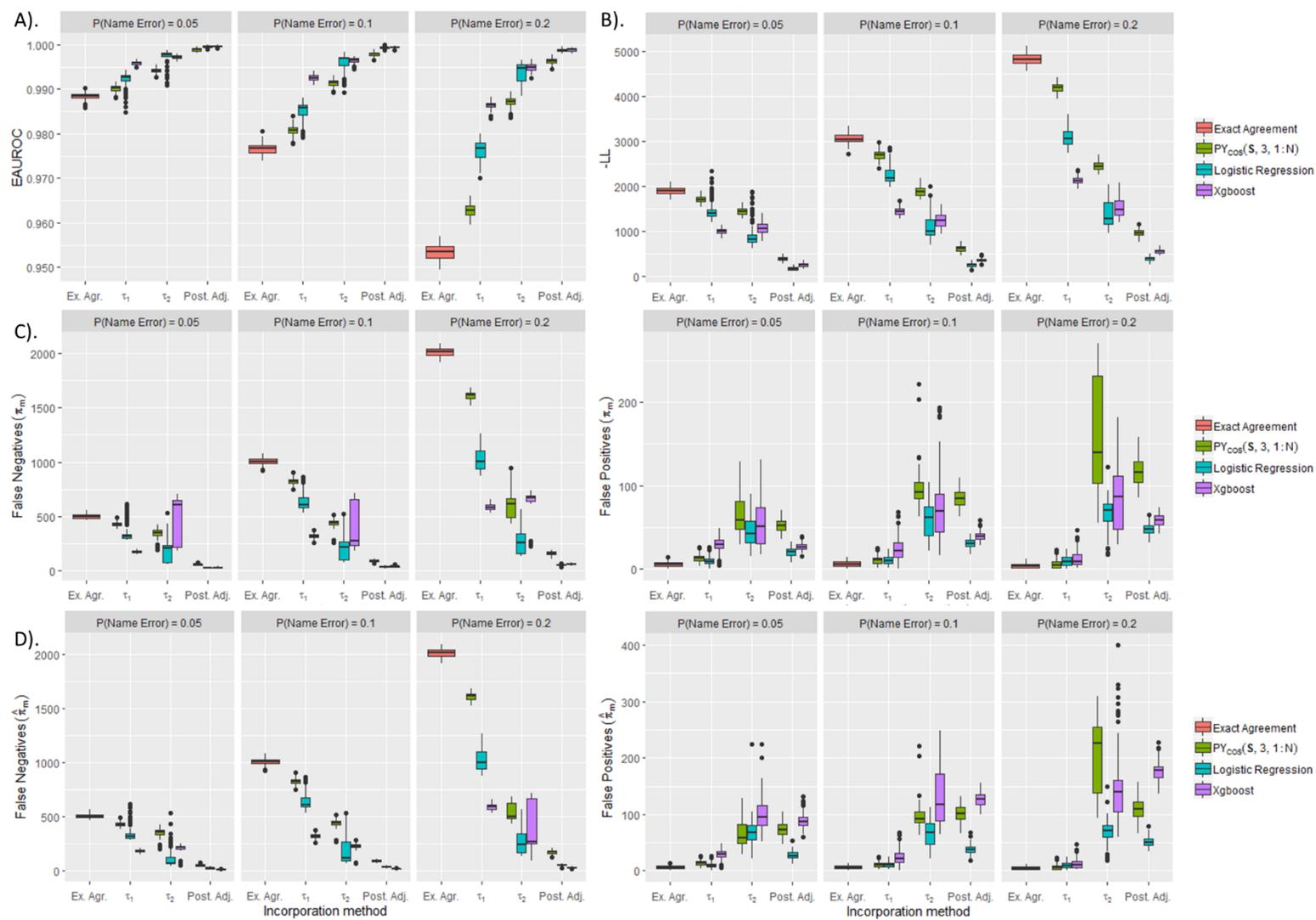

Figure 3: Linkage performance on simulated datasets with 5%, 10%, and 20% name error rates, when name, sex, year, month, and day of birth are available using various name-matching classifiers and methods of incorporation. Results are grouped by A). ranking performance (area under early portion of ROC curve); B). quality of estimated match probabilities (negative log likelihood given true match distribution); C). misclassifications when accepting the top ranked $\pi_m$ proportion of record pairs; D). misclassifications when accepting the estimated top ranked $\hat{\pi}_m$ proportion of record pairs. Each box plot corresponds to 100 simulations.

When only name, sex, and year of birth were available as identifiers, incorporation of name-matching classifiers via the $\tau_1$ and $\tau_2$ thresholding methods did not exhibit reliable improvements in performance (Figure 4). In particular, recoveries associated with the $\tau_2$ thresholding method were of very unstable quality, which further suggests the sensitivity of the method to inaccurate initial estimates of match probabilities $\zeta_j$. Additionally, performance implementing the Xgboost classifier under either thresholding method was highly variable, which could be due to differences in the distribution of scores among matches and nonmatches in the development and simulated data, as well as the tendency of boosted classification tree ensembles to push probability estimates towards 0 and 1, which may increase the sensitivity of optimal threshold score estimates to noisy inputs by flattening the predicted likelihood ratio of matches to nonmatches over much of the available range. The posterior adjustment method retained advantages over exact-agreement linkage in this setting, with consistently higher ranking performance, likelihood of estimated probabilities, and decreased misclassification rates when accepting the top true $\pi_m$ proportion of record pairs as matches. Posterior adjustment using the logistic regression classifier reduced misclassifications by 45.55% among data with 5% name error rates from 630 (568-705) to 343 (247-447); by 61.93% among data with 10% name error rates from 1127 (1043-1205) to 429 (344-514); and by 74.53% among data with 20% name error rates from 2120 (2029-2203) to 540 (465-622). However, high upper false positive rates when accepting the top estimated $\hat{\pi}_m$ proportion or record pairs indicate high variance of the overall estimated matching proportion in this setting.

Figure 4: Linkage performance on simulated datasets with 5%, 10%, and 20% name error rates, when name, sex, and year of birth are available using various name-matching classifiers and methods of incorporation. Results are grouped by A). ranking performance (area under early portion of ROC curve); B). quality of estimated match probabilities (negative log likelihood given true match distribution); C). misclassifications when accepting the top ranked $\pi_m$ proportion of record pairs; D). misclassifications when accepting the estimated top ranked $\hat{\pi}_m$ proportion of record pairs. Each box plot corresponds to 100 simulations.

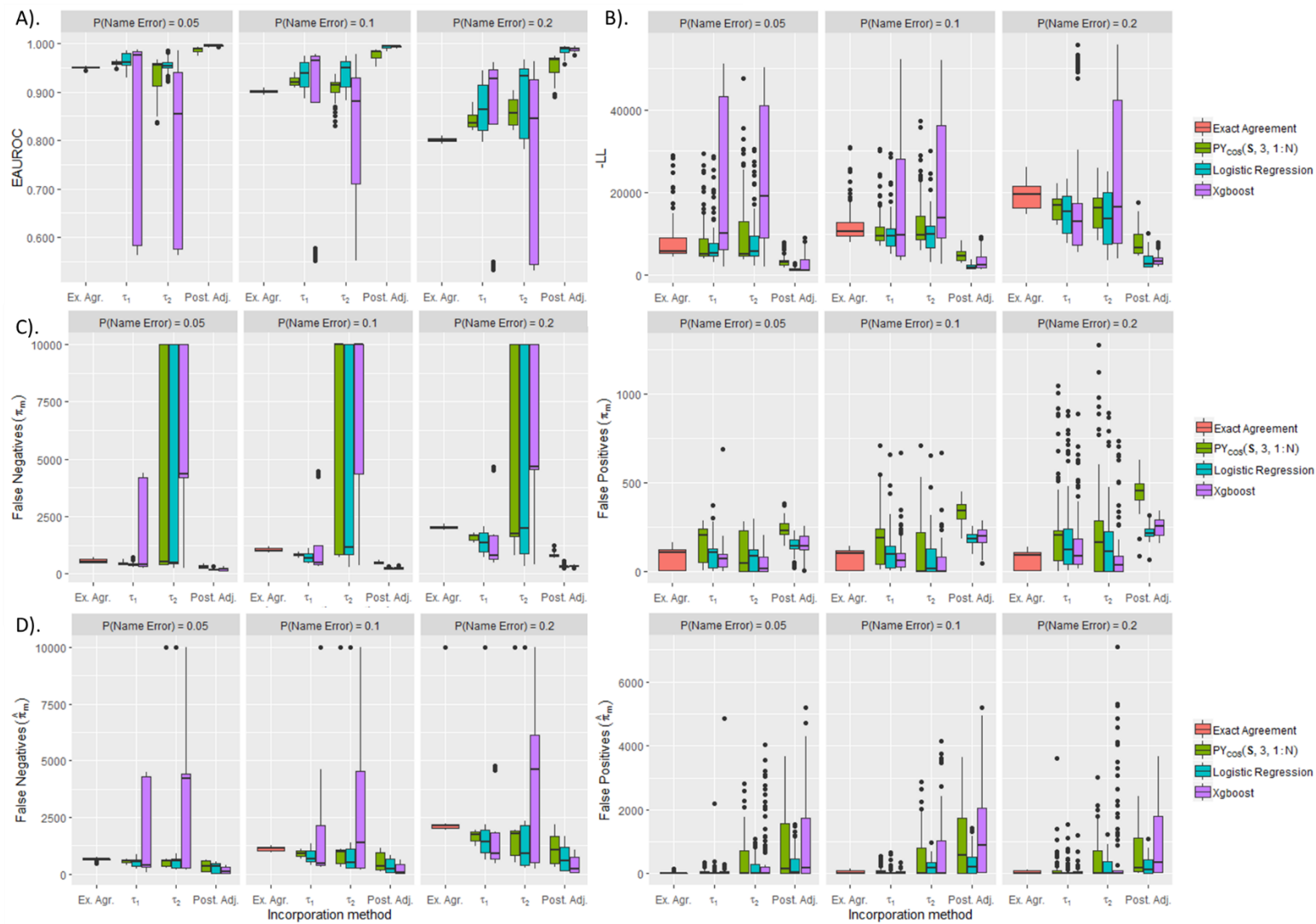

*3.5 Scalability of linkage methods incorporating name-matching classifiers*

Mean runtimes in minutes per processor per record comparison across threshold-based linkage methods were $4.17 \times 10^{-8}$ (standard deviation $2.32 \times 10^{-8}$) for exact-agreement linkage, $5.41 \times 10^{-7}$ ($6.68 \times 10^{-7}$) for the single pairwise similarity score $PY_{COS}(\mathbf{S}, 3, 1:N)$, $1.46 \times 10^{-6}$ ($6.59 \times 10^{-7}$) for the logistic regression classifier, and $2.01 \times 10^{-6}$ ($8.58 \times 10^{-7}$) for the Xgboost classifier. Runtimes per processor per record comparison were slower for the posterior adjustment method as currently implemented, due to the difficulty of implementing some of *fastLink*'s inbuilt efficient computation techniques when a limited number of record pairs are to be evaluated for name similarity: $2.38 \times 10^{-5}$ minutes ($1.59 \times 10^{-5}$) for $PY_{COS}(\mathbf{S}, 3, 1:N)$; $4.08 \times 10^{-5}$ minutes ($2.41 \times 10^{-5}$) for logistic regression; and $8.79 \times 10^{-5}$ minutes ($4.12 \times 10^{-5}$) for the Xgboost classifier. However, in most cases the number of records compared for the posterior adjustment method was a small fraction of the total record pairs: 8.35% on average (standard deviation 19%) for the single similarity score; 10.90% (21.10%) for logistic regression; and 10.40% (20.70%) for the Xgboost classifier. In practice, this ratio could be substantially reduced through user intervention to avoid comparisons among agreement patterns where the expected number of matching record pairs is very low relative to the computational burden.

## 4. Discussion

In this study, we set out to develop and implement improved methods for record linkage of datasets with name fields encoded in logographic character sets. At the outset, we hypothesized that the probability that two names are co-referent would be related to their similarity along multiple dimensions, including their pronunciation, their visual appearance, and the sequence of keystrokes required to enter them into electronic records, and that incorporating multiple predictors relating to these properties would improve data linkage. These hypotheses are supported by our results, which showed marked improvements in the ability of logistic regression and boosted classification tree classifiers incorporating diverse similarity features to separate co-referent from non-co-referent name pairs among a large sample of records. These gains in classification performance extended beyond the datasets used to train and tune the models to simulated data not drawn from the same generative distribution, suggesting at least some generalizability. The logistic regression model performed best overall, potentially due to its direct incorporation of domain knowledge in allowing the effects of individual similarity measures to vary across categories of names, as well as the tendency of generalized linear models to require less training data for reasonable predictive performance than highly data-adaptive classifiers such as Xgboost.

Optimally incorporating continuous string similarity scores into large scale record linkage is a challenging task, with some level of discretization usually implemented to overcome storage and computational hurdles. Most existing literature on the subject provides little more than rules of thumb for selecting one or more similarity thresholds to indicate complete or partial agreement of pairs of values, developed in the context of alphabetic languages (7–9). A secondary contribution of our study was to propose and test three methods for applying name-matching classifiers within, or alongside, an extended Fellegi-Sunter probabilistic record linkage framework: two were threshold-based, and one a Bayesian adjustment of prior matching probabilities based on name-matching scores. All three methods leverage initial match probability estimates based on exact-agreement linkage, as well as empirical distributions of classification-scores associated with matches and non-matches in a labeled dataset, assumed to be transferable to the new data, to allow name-matching implementations to respond to the estimated imbalance between match and non-match classes within or across patterns of agreement in identifying fields.

In linkage experiments on real and simulated datasets, the Bayesian adjustment method consistently provided the best improvements in early ranking performance, estimated match probabilities, and misclassification rates when accepting a fixed proportion of the top-ranked record pairs as matches. The apparent robustness of the Bayesian adjustment method may reflect the benefits of retaining continuous similarity information on a highly discriminating identifier, directly incorporating prior match prevalence information for each agreement pattern into posterior estimates, and adding a supervised element to the prediction process by using previously estimated probability distributions of similarity scores among matches and non-matches. On the other hand, posterior Bayesian adjustment appeared prone to overestimating the total proportion of matching record pairs. Practitioners using this method may need to carefully monitor the quality of proposed matches as they determine the cutoff match probability for declaring linkages.

The second thresholding method ($\tau_2$), which is based on optimizing the theoretical separation of matching and non-matching record pairs by name agreement, provided the next best performance when initial estimates of match probabilities were well-estimated, but exhibited instability and degraded performance when initial estimates were noisy due to reduced availability of identifying fields. The sensitivity of the $\tau_2$ thresholding method to noise in initial match probability estimates may arise from its rather naïve assumption that changes in the frequency of agreement on name will not adversely impact parameter estimation in the downstream EM algorithm.

To our knowledge, ours is the first study to incorporate methods for matching logographic names that simultaneously consider measures of phonetic, visual, and keystroke similarity into a large-scale record linkage framework. Peng et al. previously demonstrated improvement to named entity clustering in traditional and simplified Chinese using support vector machine classifiers trained on what they term character pinyin similarity features (tokens for the pinyin of single Chinese logograms) and aligned n-grams, but did not consider features related to visual similarity (17). Outside of the record linkage context, there has been substantial development on the identification of similar Chinese characters for natural language processing and educational purposes. For example, Ming et al. trained recommender systems to return similar characters on the basis of structure, semantic radical, stroke sequence, pinyin, and semantic features (26). While we did not assess these features in the current study, they may offer advantages, as could features based on image analysis of character inputs (e.g., (27)). In using string similarity scores to adjust parameters previously estimated via Fellegi-Sunter probabilistic linkage, our posterior Bayesian adjustment method is somewhat similar to a method proposed by Li et al. to use linearly combined string similarity scores across multiple identifying fields, weighted by the log likelihood ratio of their exact agreement pattern among matches and nonmatches (a Fellegi-Sunter model output) (28). However, our approach is better justified on the basis of probability theory, and allows for conditional dependence between field agreements to inform the initial estimation of match probabilities.

While the performance and scalability of the methods we have developed for linkage with logographic name-matching are encouraging, it is not clear how well the classifiers trained to our dataset will generalize to new data in different contexts. Nonetheless, we expect our framework of assessing and incorporating comparisons of multiple representations to capture and weight phonetic, visual, and keystroke similarities in machine learning classifiers will provide good performance in general if adequately adapted to specific contexts. There is substantial room for improvement in the particular feature encodings we used, which were drawn for convenience from publicly available datasets. Improvements could include incorporation of direct phonetic encodings, enhanced measures of visual similarity, and more carefully-selected features for radical decompositions. Cognitive linguistic research has revealed that the structure of compound characters, as well as semantic radicals (the element of the character which indicates its domain of reference) are important factors in perceptual character similarity (29), and we noted that most, if not all, instances of a character being decomposed into its radicals corresponded with a vertical bisection. While they generally demonstrated improved performance over an exact-agreement baseline, and provide some theoretical justification to tune the application of name-matching classifiers to specific record linkage problems, none of our proposed methods for incorporating name-matching into linkage is entirely satisfying: The threshold-based methods lose substantial information by dichotomizing continuous similarity scores, while the Bayesian posterior adjustment method assumes that name similarity is invariant with respect to agreement on other fields, and lacks flexibility to adapt the estimated probability of matching given observed string similarity scores to new data. An alternative approach may be to treat disjoint intervals of string similarity as levels of agreement within the linkage mixture model, as proposed by Winkler and others (9,30,31). Such an approach requires more parameters to be estimated during probabilistic linkage, and complicates the modeling of conditional dependence between fields. However, both issues may be mitigated by using GAMs with suitable regularization, such as monotonicity constraints. We leave this and other improvements to future studies.